\documentclass{jaa}
\usepackage{natbib}
\usepackage{graphicx}

\begin{document}\sloppy


\title{Performance of the UVIT Level-2 Pipeline}



\author{S. K. Ghosh\textsuperscript{1,5,*}, S. N. Tandon\textsuperscript{2,3},
 P. Joseph\textsuperscript{3}, A. Devaraj\textsuperscript{3},
 D. S. Shelat\textsuperscript{4,5}, \and C. S. Stalin\textsuperscript{3}}
\affilOne{\textsuperscript{1}Tata Institute of Fundamental Research, Mumbai 400005, India\\}
\affilTwo{\textsuperscript{2}Inter-University Centre for Astronomy \& Astrophysics, Pune 411007, India\\}
\affilThree{\textsuperscript{3}Indian Institute of Astrophysics, Bangalore 560034, India\\}
\affilFour{\textsuperscript{4}Space Application Centre (ISRO), Ahmedabad 380015, India\\}
\affilFive{\textsuperscript{5}National Centre for Radio-Astrophysics (NCRA-TIFR), Pune 411007, India\\}



\twocolumn[{

\maketitle

\corres{swarna@tifr.res.in}


\begin{abstract}
Performance of the Level-2 pipeline, which translates the UVIT data created by
 the ISRO's ground segment processing systems (Level-1) into astronomer ready
 scientific data products, is described. This pipeline has evolved significantly
 from experiences during the in orbit mission. With time, the detector
 modules of UVIT developed certain defects which led to occasional corruption
 of imaging and timing
data. This article will describe the improvements and mitigation plans 
incorporated in the pipeline and
report its efficacy and quantify the performance.

\end{abstract}



}]
\msinfo{ }{ }



\section{Introduction}

The Ultra-Violet Imaging Telescope (UVIT) on board AstroSat carries out simultaneous imaging
of the sky in three bands, viz., Far-UV (FUV : 130 -- 180 nm), Near-UV (NUV : 200 -- 300 nm) \&
visible (VIS : 320 -- 550 nm). Using a selectable filter for each band, radiation in limited ranges of
wavelength are allowed to reach the respective detectors. Gratings for UV bands also allow low
spectral resolution slit-less spectroscopy. The detector for each of the 3 bands consist of an image
intensifier assembly consisting of a photo-cathode deposited on the window which proximity focuses
photo-electrons onto a pair of Micro-Channel Plate assemblies biased to suitable high voltages to
multiply them. These secondary electrons are further accelerated by electric field to hit a phosphor
(acting as anode) to generate corresponding optical light pulses. This light is coupled through a de
magnifying fibre-optic taper onto a CMOS imager (512x512 pixels) which covers the entire circular
active area of the detector. The full field of view of UVIT is 28 arc-min (dia) for each band and it is
read at a rate $\sim$ 28.7 frames/s, but a centrally positioned square window of selectable size allows for
imaging in smaller fields (smallest : 5.5${}^{\prime} \times 5.5{}^{\prime}$ corresponding to 100x100 pixels) with proportionately
faster sampling in photon counting mode imaging operation of the UV detectors. The Far-UV and
Near-UV bands are operated in photon-counting mode, with a high gain of the Micro-Channel-Plate
assembly, while the VIS band is operated in an integration mode, with a low gain of the Micro-Channel
Plate assembly and a much slower image read out rate which aid determination of drift of the
spacecraft with high accuracy. Each band is configured for sky observations by effecting several user
selectable parameters, the major ones being – filter / grating \& window size (coupled image frame
rate). Further details about UVIT can be found elsewhere (Kumar et al. ~2012a, Tandon et al. ~2017a,
Tandon et al. ~2017b).

The data originating from UVIT (detector modules /electronics, filter drive units) as well as the
spacecraft systems (time, aspect, etc) are sorted in time and collated on ground by various processing
stages at the centres of ISRO (ISTRAC, URSC, ISSDC). The resulting data products, called Level-1
(L1), are provided to Payload Operation Centre, POC, for UVIT (at IIA, Bangalore). At POC further
processing are carried out using the UVIT’s Level-2 pipeline along with some auxiliary programs,
which result in final astronomer ready L2 products for dissemination and archiving. The most
significant role of the Level-2 pipeline is to first extract spacecraft drift parameters with time (from
stars detected in sky images in VIS) and apply corresponding corrections, along with various
systematic effects inherent within UVIT, to combine the series of short exposure UV data into
integrated sky images.

The development of the UVIT’s Level-2 pipeline, hereafter ``pipeline" was initiated well before the
launch of AstroSat / UVIT, but significant changes of strategy / algorithms / coding were needed to
be carried out after real L1 data corresponding to in orbit operations became available. This article
chronicles the evolution of the pipeline over the initial years of after the launch in September 2015. The
performance of the pipeline in terms of achieved angular resolution (characteristics of Point Spread
Function) and efficiency with which input L1 data could be translated into L2 products have been
quantified.

\section{Functionalities of the pipeline}

Typical observation of any astronomical target is organized as a sequence of exposures in selected
Filter-Window size configurations of Far-UV \& Near-UV bands as per scientific requirements of the
observer. The VIS band is configured with largest window (512x512; covering the full 28 arc-min sky field)
and a `safe' filter such that the brightest star in the field would not cross the nominal safety threshold
set to trigger the on-board autonomous Bright Object Detect, BOD, logic of the detector. Identical
detector safety concerns are also addressed while selecting acceptable filters for the UV bands. The
UVIT can observe only during the dark part of any orbit due to large background from scattered solar
and earth radiation in the bright part. This allows a maximum of $\sim$ 2000 second uninterrupted exposure
at one instance (this could become even shorter in practice, due to passage of spacecraft through South
Atlantic Anomaly, SAA, or constraints like minimum sun angle, etc). Hence, observation needing
longer total integration time is spread over multiple orbits. Each uninterrupted imaging operation with a
fixed Filter-Window configuration is called an ``Episode". The dark part of one orbit can also
accommodate more than one Episode (e.g. shorter exposures with different Filter / Window). Generally,
observations of a specific target are scheduled in successive orbits sandwiched between two spacecraft
slews – one for pointing to the target and the next away from it (typically pointing to another target)
called a ``Pointing". Accordingly, UVIT observations from each Pointing would generally contain
multiple Episodes. The complete Level-1 data from a particular Pointing are finally combined by the
ISRO ground station software systems into a single ``Merged L1" data bundle. Initially, ISRO promptly
provides Level-1 data from individual orbit dumps (say, “Orbit L1”). While Merged L1 would
mandatorily hold data from multi-Episode observations for each band of UVIT, the Orbit L1 would
most often hold single Episode per band. The L1 data bundle for UVIT consists of Science (Imaging)
data from all 3 Detectors and Auxilliary (Aux) data containing inputs regarding UVIT Filters and
various information from Spacecraft Systems, e.g. time calibration, attitude of satellite reference axes,
position in orbit, gyro sensors' output \& house keeping information.

The pipeline has been designed to automatically handle an entire input L1 data bundle at a time and
carry out a series of tasks sequentially to generate all output products from its single run.
 It uses the UVIT's Calibration Database (CALDB) and the user selected various parameters \& switches through a
Parameter Interface Library (PIL).

As mentioned in the introduction,
the detectors for all 3 bands of UVIT are constructed identical intensified CMOS-imaging systems,
which can be operated either with a high gain effecting Photon Counting mode (PC) or at a low gain
effectively functioning in Integration mode (INT). The detectors for Far-UV \& Near-UV bands are
operated in PC mode and frames are read out at a high speed ($\sim$ 28.7 fps for full field; max $\sim$  640 fps
for smallest field) while for VIS band INT mode is used at a low frame rate ($\sim$ 1 fps). Successive
frames from the CMOS imager are read out continuously during active imaging (Kumar et al. ~2012b,
Tandon et al. ~2017c, Tandon et al. ~2020). The on board processing of these frames depends on the mode.
For INT mode, pixel values for the image within the selected Window constitute the detector data. On
the other hand, for PC mode, for each frame the part within the selected Window is processed to
identify photon events from the light distribution and compute their centroid coordinates along 2 axes
of the detector. The details of individual detected photon events constitute the imaging data for PC
mode. In view of the above mentioned differences in the data emanating from the detector operated in
INT / PC modes, the processing schemes employed in the pipeline are separately dedicated for each
mode.

The key functionality of the pipeline is to translate UVIT measurements to astronomer ready sky
images by recovering absolute aspect and the angular resolution in presence of various systematic
effects and random perturbations of the spacecraft pointing. While the many processes involved to
achieve it are rather complex, they are divided into two main themes : (a) extraction of the drift and
disturbances to the optical axes of UVIT / spacecraft reference axes, and use these to combine the
frames using shift and add algorithm \& (b) generation of sky Images of Intensity \& corresponding
Exposure Arrays and Uncertainty Arrays, by applying all applicable corrections for instrumental
effects, e.g. flat-field corrections and corrections for distortion as quantified in the Calibration
Database. Exposure Arrays are required because exposure over the field varies due to drift of the
pointing, and Uncertainty Arrays are required because the actual number of photons depends on flat
field corrections in addition to the exposure and the intensity. Each of these have been implemented as
stand alone processing chain handling UVIT single band data from one Episode of observation, named
– Relative Aspect (RA) \& Level-2 Imaging (L2) respectively. Given that two very different modes of
operation of UVIT, viz., INT \& PC, each chain gets further diversified into two, making in total four
chains. The relevant chains are operated on individual Episode data sequentially since the L2 chain
needs results from RA chain run for corresponding time range. Most common instance of processing of
data from a specific Episode, involves execution of the RA chain on VIS band data to extract drift
series followed by two separate executions of the L2 chain on NUV \& FUV band data using the
corresponding drift series. In the rare situation of absence of VIS data for the particular Episode, drift
series is extracted using the NUV data. This sequence completes processing for one Episode generating
a complete set of resulting products. The most important ones : UV sky Images of Intensity, Exposure
\& statistical Uncertainty arrays (4800x4800; pixel $\sim$ 0.4 arc-sec) in both the Detector as well as
Equatorial coordinate systems (along with Astrometric corrections when successful), final corrected
UV event table with details for timing studies.

As stated earlier, in general the total exposure for a target with a specific Filter-Window
configuration for an UV band would constitute multiple Episodes. Hence, additional processing are
involved in combining the results from individual Episodes. The ``combining" operation of multiple
Episodes involves determination relative shifts \& rotations between UV images from individual
Episodes and applying these corrections to align and then stacking. This is carried out in the following
steps : identification of a `master' Episode (with largest exposure); tabulation of stars \& their centroidcoordinates detected in the UV Intensity image from each Episode; correlation of stars between
`master' \& every other Episodes and determine shifts \& rotations relative to the `master'; application of
shifts \& rotations to products of all non-master Episodes aligning then with the `master' Episode;
accumulation of corresponding products from all Episodes generating the final ``combined" products.
One important detail regarding generation of combined UV Intensity image -- the offsets for each
Episode are applied to the centroids of individual photons first and finally gridding them on to a
common array, thereby retaining the precision / angular resolution achieved in individual Episodes,
even in the most general case for offset including a rotation in addition to shifts along the two axes. On
the other hand, all shift \& rotation operations for Exposure \& Uncertainty arrays are performed on the
array elements (following first sub-division to 9600x9600 then shift \& rotation operations on sub
pixels, re-gridding followed by binning back to 4800x4800 to minimize the loss of precision due to
finite pixel size). The success of this scheme for combining critically depends on availability of bright
enough UV stars in the field. Accordingly, the success rate for the FUV band is lower than that for the
NUV band. The astrometric refinements are carried out on the combined multi-Episode UV images
also. All the usual L2 products (except events list) are generated for the multi-Episode cases.

\section{Evolution of the pipeline during early in orbit operations of UVIT}

The development of the pipeline had begun well before the AstroSat /UVIT launch when the
instrument details were frozen after Engineering Model was realized successfully. The architectural and
structural detail of the pipeline were finalized early on. Given the similarities of certain processes used
in multiple chains, these were developed as modules. The chains called several such modules in
appropriate sequences supplemented by unique processes dedicated for the chain. Once the laboratory
data from UVIT detectors became available, many functionalities of the modules could be tested and
validated. However, the real inputs for the pipeline, the L1 data bundles, became available only after
UVIT turned on was in orbit. For the first time concurrent data originating from the spacecraft sub
systems along with UVIT could enable testing of modules dependent on such interconnected dataset.
Initially there were some surprises leading to discovery of minor oversights in the pipeline realization.
This resulted in many additional features to be incorporated. Here we briefly describe a few of them.

It may be recalled that the data from the UVIT Filter drive units are collected directly by the
spacecraft sub-system appearing in the Aux data in L1 as opposed to UVIT Detector data from UVIT
electronics as Science data stream. The L1 product introduces the Filter information in to Detector data
using time correlation between Aux \& Science data. However, it was noticed that the reported identity
of the Filter was incorrect at times. Since no credible reason could be attributed to this anomaly, UVIT
POC devised a scheme to correct for his and generating a L1${}^{\prime}$ data bundle, which was sent to ISSDC
for archiving and dissemination.

\begin{figure} [th]
\includegraphics[scale=0.35]{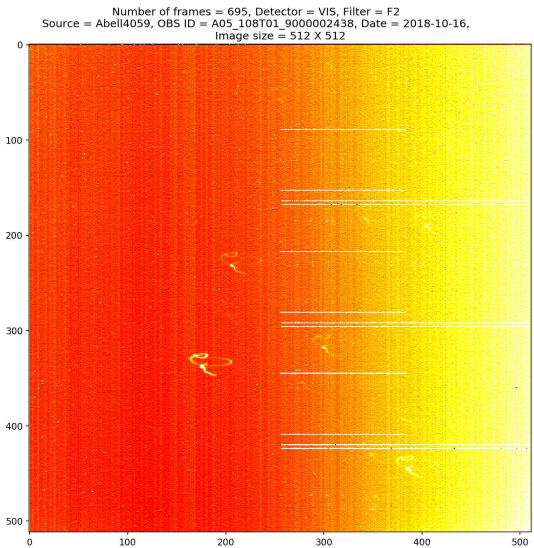}
\caption{Example of stripes (along horizontal axis) in Quick
Look image generated from VIS band images from one
Episode. Such artifacts are handled successfully
by the Pipeline while extracting the drifts. A few curved
tracks correspond to detected stars and their movement
due to spacecraft drift during the observation Episode.}

\end{figure}

As time progressed, the UVIT detector electronics was found to vulnerable to charged particle hits
(Single Event Effects / Latchups). This manifested as peculiar anomalies in the Science data stream
rendering many logics of the pipeline to fail. A few examples are described here. The raw images of
VIS band showed bright (pixel with higher ADU counts) stripes along one axis of the CMOS images,
which affected the success of star finding algorithm. Since the affected pixels appeared in a pattern and
also constituted a negligible fraction, appropriate logic was developed and incorporated in the pipeline
to address them without any loss of functionality (introducing a new block named ``Artifact Handler").
As time progressed, the number of such stripes increased (see Fig 1.), which required further tweaking
of the fix. Later, the positions of the stripes too started jumping around, which required completely new
kind of logic to by pass them. Still, the RA\_INT chain could successfully extract the drift series
without any degradation of accuracy.
Since mid-2017, no new kind of artifacts have been observed in the VIS band images and
the current Artifact Handler
 continues to mitigate such effects successfully. 


 At times the Science data from successive Episodes for the NUVband showed extremely unusual patches rendering these dataset to be completely unusable (see Fig. 2).
Fortunately, a power RESET could recover the NUV band from these artifacts.

\begin{figure} [th]
\centering
\includegraphics[scale=0.35]{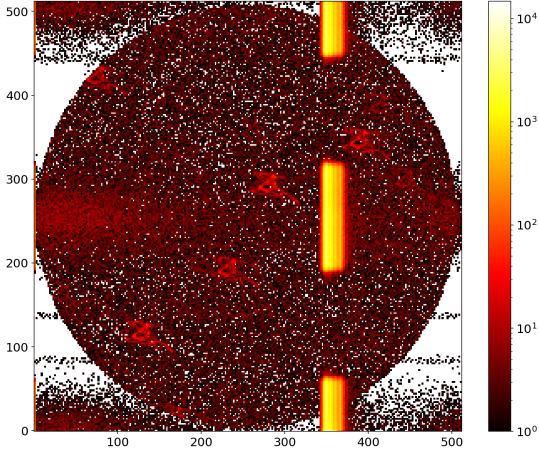}
\caption{
Quick Look image in NUV band generated by plotting
centroid coordinates of all photons detected during one observation
Episode. The broad patches \& narrow stripes (along
horizontal axis) are unexpected artifacts.
}

\end{figure}

The L1 dataset was expected to correlate UVIT's internal clocks (one per band) to well calibrated
Universal Time Clock (UTC) based on periodic simultaneous time sampling by the spacecraft. Hence,
the pipeline was originally designed using UTC as the primary reference for time. However, often this
time correlation was found to be unreliable. Accordingly an additional functionality was introduced in
the pipeline which allows use of UVIT's Master Clock for timing (UVIT is configured selecting an
unique band, VIS, as Master for all the 3 bands ensuring inter-band time synchronization). This timing
scheme has since been used successfully. Even, while by passing the UTC for primary processes in the
pipeline, a parallel scheme has been introduced to provide approximate (good to $\sim$ 1 sec) absolute time
(MJD\_UT) for every frame, by identifying selected patches of time where UTC correlates well with the
UVIT Master Clock. This allows timing studies using the pipeline product providing photon event table
with all details including MJD\_UT, even in UTC by passed mode.

In the L1 Science data for UV bands operated in PC mode, various anomalies were noticed. For
example : jumps (spike) in Frame Number or Frame Time, completely discontinuous Frame Number
\& Frame Time inserted within normal good data sequence; data from a particular frame repeated
elsewhere in the data from the same Episode; abrupt discontinuity in Frame Number which violating
monotonicity. These artifacts were discovered gradually during early in orbit phase. Appropriate
remedial logic were designed and incorporated within the DataIngest block of the pipeline.

\begin{figure} [th]
\centering
\includegraphics[scale=0.30]{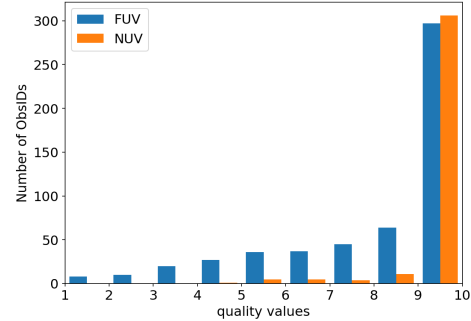}
\caption{
Distribution of the quality values (from achieved PSF
size and amount of pedestal) for FUV and NUV band images.
}

\end{figure}

\section{Performance of the Pipeline}

The Pipeline is routinely operated at the UVIT Payload Operation Centre (POC) at the Indian
Institute of Astrophysics, Bangalore. The L1 data are regularly received by the POC from ISSDC
/ISRO. The POC sends L1${}^{\prime}$ data (modified) and L2 product bundles back to ISSDC for dissemination
to the Proposers and archiving.
An installer for the Pipeline, Calibration Database and relevant instructions
are publicly available from the sites :
https://uvit.iiap.res.in/Downloads; 
http://astrosat-ssc.iucaa.in/?q=uvitData; 
https://www.tifr.res.in/{\char`\~}uvit/.

Though the pipeline has undergone major evolution during the initial year or so, it has since
stabilized and has been performing satisfactorily. The data for the early phase are also being re
processed using the last stable version at the POC.


The astrometric accuracy achieved over the central 24$^{\prime}$ diameter of
the field is  better than $0.3^{{\prime}{\prime}}$ (rms), but it is not so good for larger diameters
 and the errors could be up to $0.8^{{\prime}{\prime}}$ for parts of the outer annular region
with radius between 12$^{\prime}$ and  14$^{\prime}$. The photometric error, as determined from
 multiple observations of SMC fields, is
found to be $<$6\% (rms) (Tandon et al. 2020).


Two key parameters to quantify the success of the pipeline are : (a) size of the PSF and (b) the
amount of L1 data incorporated in the final L2 sky products. The former checks not only the quality of
drift tracking but also application of all corrections for systematic effects. Accordingly, the POC has
devised a quality factor which grades the intensity image products from the FWHM size of the core of
the PSF as well as the fraction of intensity in the pedestal, for a few point like sources spread across the
field. The best value of the quality factor is `10' gradually reducing with degradation to lower values.
Fig. 3 shows the distribution of the quality factor for a large sample of pipeline products. The quality
report for every data set is always included in the standard L2 product bundle. For a sizable fraction of
the observations in both NUV \& FUV bands, a score of 9 or above has been achieved. This implies the
FWHM of the PSF to be $<$ 1.8 arc-sec \& the pedestal $<$ 20\%.

\begin{figure} [th]
\centering
\includegraphics[scale=0.30]{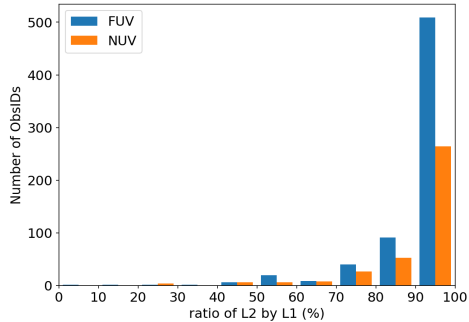}
\caption{
Distribution of the yield (fraction of input Level-1 data
translated to Level-2 image products, for FUV and NUV bands)
achieved by the pipeline.
}

\end{figure}

The second aspect relates to the ``yield" of the pipeline. The sum of exposure time of frames
contributing to the final sky images (T\_L2) is compared with the corresponding 
frames available in the
L1 data set (T\_L1), as a fraction, yield = (T\_L2/T\_L1). The distribution of yield among a large sample
of datasets is presented in Fig. 4.

\section{Future plans}

The UVIT's Level-2 pipeline has been performing quite satisfactorily as evidenced in the earlier
section. However, at times some issues have hampered the generation of Merged L1 data sets. Since the
L1 datasets corresponding to individual dump orbits are available, a scheme to stitch them in to pseudo
merged L1 data has been developed at the POC. It has been planned to generate the missing merged L1
data using this scheme. Another plan for future augmentation of the pipeline is to utilize optical stars
detected in VIS frames in the logic for combining multi-Episode data (instead of UV stars used
currently). This would also help improving the success rate of astrometry block.

\section*{Acknowledgements}


The UVIT project is a result of collaboration between IIA, Bangalore, IUCAA, Pune, TIFR, Mumbai,
many centers of the Indian Space Research Organization (ISRO), and the Canadian Space Agency. We
thank these organizations for their support. We gratefully thank members of the Ground Segment
software teams of ISRO for their support. We also thank members of the AstroSat Project \& the
AstroSat Science Working Group for their feedback.

\vspace{-1em}



%
\end{document}